\documentclass[final]{cvpr}
\usepackage{times}
\usepackage{epsfig}
\usepackage{graphicx}
\usepackage{amsmath}
\usepackage{amssymb}
\usepackage{multirow}
\usepackage{graphicx}
\usepackage{tabularx}
\usepackage{xcolor}
\usepackage{caption}
\usepackage{subcaption}
\usepackage{array}
\usepackage{url}
\usepackage{algorithm}
\usepackage{algpseudocode}
\usepackage{appendix}
\usepackage[pagebackref=true,breaklinks=true,colorlinks,bookmarks=false]{hyperref}

\begin{document}

\title{Slimmable Compressive Autoencoders for Practical Neural Image Compression}

\author{Fei Yang$^{1,2}$, Luis Herranz$^{2}$, Yongmei Cheng$^1$, Mikhail G. Mozerov$^{2}$\\
$^{1}$ School of Automation, Northwestern Polytechnical University, Xi'an, China\\
$^{2}$ Computer Vision Center, Universitat Autonoma de Barcelona, Barcelona, Spain\\
{\tt\small \{fyang,lherranz,mozerov\}@cvc.uab.es, chengym@nwpu.edu.cn}
}
\maketitle

\begin{abstract}
    Neural image compression leverages deep neural networks to outperform traditional image codecs in rate-distortion performance. However, the resulting models are also heavy, computationally demanding and generally optimized for a single rate, limiting their practical use.	
	Focusing on practical image compression, we propose slimmable compressive autoencoders (SlimCAEs), where rate (R) and distortion (D) are jointly optimized for different capacities. Once trained, encoders and decoders can be executed at different capacities, leading to different rates and complexities. We show that a successful implementation of SlimCAEs requires suitable capacity-specific RD tradeoffs. Our experiments show that SlimCAEs are highly flexible models that provide excellent rate-distortion performance, variable rate, and dynamic adjustment of memory, computational cost and latency, thus addressing the main requirements of practical image compression.
\end{abstract}

\section{Introduction}
	Visual information (e.g. images, videos) plays a central role in human content creation, communication and interaction. Efficient storage and transmission through constrained channels requires compression. We can thus consider the \textit{basic (lossy) image compression} problem, where the goal is to obtain the shortest  binary representation (i.e. lowest \textit{rate} bitstream) that can represent the input image at a certain level of fidelity (i.e. minimum \textit{distortion}). Thus, low rate and low distortion are fundamentally opposing objectives, in practice involving a \textit{rate-distortion (RD) tradeoff}. The more challenging problem of \textit{practical image compression} further includes real-world constraints such as memory, computation and latency, related with their implementation in resource-constrained devices (e.g. mobile phones) and networks. Similarly, video compression addresses the same problem for sequences of images, where low complexity and latency become even more critical~\cite{wiegand2011source}. Many applications also require dynamic control of the RD tradeoff to adapt to specific rate requirements (i.e. \textit{variable rate}). 

	\begin{table}[t]
		\centering
		\resizebox{\columnwidth}{!}{
			{\renewcommand{\arraystretch}{0.8}
				\setlength{\tabcolsep}{2pt}
				\begin{tabular}{ccccccc}
					\hline
					\multirow{ 2}{*}{Method} & \multirow{ 2}{*}{\shortstack{Rate-dist.\\perform.}}  & \multirow{ 2}{*}{\shortstack{Total\\memory}} & \multicolumn{3}{c}{Variable} & \multirow{ 2}{*}{\shortstack{Training\\time}}\\
					& & & Rate & Memory & FLOPs\\
					\hline 
					JPEG,JP2K & Low & Very low & Yes & - & - & -\\
					BPG & High & Low & Yes & - & - & -\\
					\hline
					Single CAE & Optimal  & Medium & No & No & No & Low \\
					Multiple CAEs & Optimal  & High & Yes & Yes & Yes & High \\
					BScale\cite{theis2017lossy} & Medium   & Medium & Yes& No & No & Low\\
					MAE\cite{yang2020variable},cAE\cite{choi2019variable} & High   & Medium & Yes& No & No & Low\\
					\hline
					SlimCAE & Optimal  & Medium & Yes& Yes & Yes & Low\\
					\hline
				\end{tabular}
			}
		}
		\vspace{-0.8em}
		\caption{Comparison of compression methods.}
		\label{tab:comparison_methods}
	\end{table}

	\begin{figure}[t]
	    \vspace{-0.8em}
		\centering
		\includegraphics[width=\columnwidth]{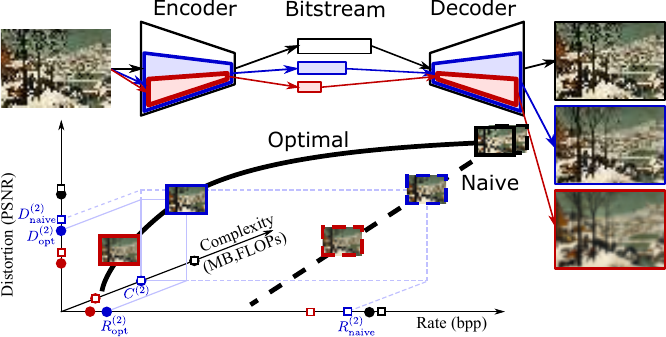}
		\caption{Variable rate and complexity adaptive image compression with a slimmable compressive autoencoder.}
		\label{fig:slimCAE}
		\vspace{-0.8em}
	\end{figure}

	Traditional methods (e.g. JPEG~\cite{wallace1992jpeg}, JPEG2000~\cite{rabbani2002jpeg2000,skodras2001jpeg}, BPG~\cite{sullivan2012overview}) follow the transform coding paradigm with carefully designed linear transforms and coding tools, to effectively address practical image compression. Encoding is block-based and iterative. RD is optimized during encoding by exhaustively searching optimal block partitions and coding and prediction modes. Complexity can be controlled by limiting the range of coding and prediction modes, or using heuristics. The rate can be estimated from previous blocks, and controlled by adjusting quantization parameters.
	
	A more recent paradigm is \textit{neural image compression} (NIC)~\cite{toderici2015variable, balle2016end,gregor2016towards,toderici2017full,theis2017lossy,johnston2018improved, liu2018cnn,mentzer2018conditional, minnen2018joint,li2018learning} (or learned image compression), which exploits flexible nonlinear transforms and entropy models parametrized as deep neural networks. The framework typically consists of an autoencoder followed by quantization and entropy coding (henceforth a \textit{compressive autoencoder} -CAE-~\cite{theis2017lossy}). While they can outperform traditional image codecs in RD, they are typically only optimal for a single target rate, and at the cost of much heavier and computationally expensive models that require specialized hardware (e.g. GPUs), making them unattractive in practical resource-limited scenarios. In contrast to block-based codecs, NIC approaches are typically image-based and feed-forward, resulting in a constant processing cost. While some approaches address variable rate~\cite{theis2017lossy,cai2018efficient,yang2020variable,choi2019variable}, practical concerns related with efficiency still remain largely unaddressed (see Table~\ref{tab:comparison_methods}).
	
	The challenge of deploying deep neural networks in resource-limited devices (e.g. smartphones, tablets) has motivated research on lightweight architectures~\cite{howard2017mobilenets,sandler2018mobilenetv2}, integer and binary networks~\cite{khan1994integer,rastegari2016xnor,jacob2018quantization} and automatic architecture search~\cite{tan2019mnasnet}. However, only few works have addressed efficiency in NIC~\cite{rippel2017real,johnston2019computationally}. We borrow the idea of slimmable neural networks~\cite{yu2019slimmable}, where the width of the layers (i.e. number of channels) of a classifier can be adjusted to trade off accuracy for computational efficiency.

In this paper, we propose the {slimmable compressive autoencoder (SlimCAE)} framework, where we show that the slimming mechanism can enable both variable rate and adaptive complexity (see Fig.~\ref{fig:slimCAE}). We propose and study different variants of slimmable generalized divisive normalization~\cite{balle2015density} (GDN) layers, and slimmable probability models.
Naive training of SlimCAEs, with the different subnetworks (i.e. \textit{subCAEs}) optimizing the same loss on all widths, results in suboptimal performance. We crucially observe that each RD tradeoff has an corresponding minimum capacity. This suggest that, in contrast to other slimmable networks, each width should have different objectives, i.e. different $D+\lambda R$, determined by the corresponding tradeoff $\lambda$. This characteristic makes SlimCAEs more difficult to train, and unlikely to benefit from implicit or explicit distillation~\cite{yu2019slimmable}. Addressing this problem, we propose $\lambda$-scheduling an algorithm that alternates between training the model and adjusting the different $\lambda$s. Via slimming, SlimCAEs can address the main requirements of \textit{practical neural image compression (PNIC)} in a simple and integrated way.

Our main contributions are: (1) a novel rate and complexity control mechanism via layer widths, motivated by a key insight connecting optimal RD tradeoffs and capacity; (2) the SlimCAE framework, which enables control of computation, memory and rate, required for PNIC; (3) an efficient training algorithm for SlimCAEs; (4) novel slimmable modules (i.e. GDNs, entropy models). In addition, SlimCAEs can be easily adapted to obtain scalable bitstreams.

\section{Related work}
\subsection{Neural image compression}
The modern non-linear deep autoencoding framework with quantization and entropy coding (which here we refer to as compressive autoencoder~\cite{theis2017lossy}), trained by backpropagation to minimize a (fixed) combination of rate and distortion, is relatively recent~\cite{toderici2015variable,theis2017lossy,balle2016end}. Encoders and decoders often integrate multi-scale structures~\cite{cai2018efficient, rippel2017real, nakanishi2018neural} and generalized divisive normalization (GDN) layers~\cite{balle2015density,balle2016end}. End-to-end training requires replacing non-differentiable quantization by differentiable proxies such as additive noise~\cite{balle2016end}, identity in the backward pass~\cite{theis2017lossy} and soft-to-hard vector quantization~\cite{agustsson2017soft}. Entropy coding also benefits from learnable CNNs, via hyperpriors~\cite{balle2018variational} and contextual models~\cite{mentzer2018conditional, lee2018context, li2020efficient, minnen2018joint, minnen2020channel}. More recently, adversarial training has been used to target very low rates~\cite{rippel2017real,tschannen2018deep,agustsson2019generative}.

\subsection{Variable rate image compression}
\begin{figure*}[!ht]
	\centering
	\includegraphics[width=0.92\textwidth]{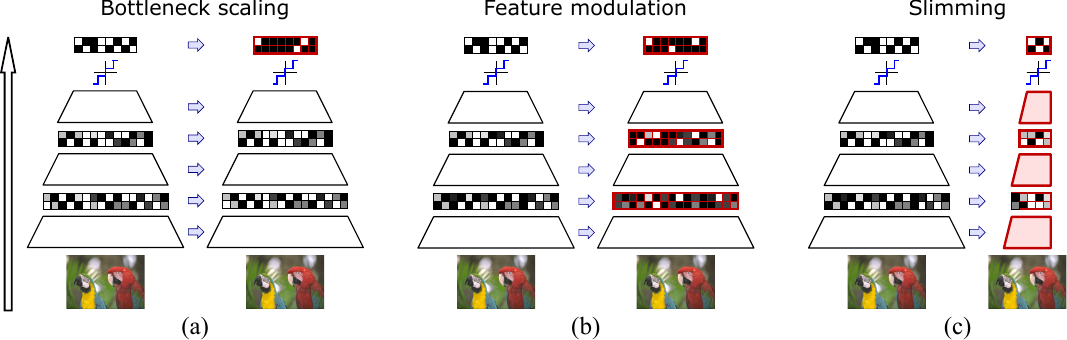}
	\vspace{-0.8em}
	\caption{Mechanisms to achieve variable rate in compressive autoencoders: (a) bottleneck scaling~\cite{theis2017lossy}, (b) feature modulation~\cite{yang2020variable,choi2019variable}, and (c) proposed method (SlimCAE). Adaptation from high rate (left) to low rate (right). Changes are highlighted in red. Only SlimCAE reduces memory and computation. GDN layers are not included for simplicity. }
	\label{comparison_baselines}
	\vspace{-1.0em}
\end{figure*}

Many practical applications require certain control of the target rate. Traditional image compression methods enable this functionality via quantization tables that scale DCT coefficients according to the target rate. Similar to traditional methods, Theis \textit{et al.}~\cite{theis2017lossy} learns a set of rate-specific parameters to scale the bottleneck feature before quantization (we refer to this as \textit{bottleneck scaling}, see Fig.~\ref{comparison_baselines}a). Modulated autoencoders (MAEs)~\cite{yang2020variable} and conditional autoencoders (cAEs)~\cite{choi2019variable} show that modulating also intermediate features improves RD performance (see Fig.~\ref{comparison_baselines}b).  
Recurrent neural networks can also realize variable rate coding~\cite{toderici2015variable}, yet are demanding computationally. Cai \textit{et al.}~\cite{cai2018efficient} proposed a multi-scale decomposition network, each scale targeting a different rate. None of these methods provides explicit control over complexity. Our approach (see Fig.~\ref{comparison_baselines}c), in contrast, can jointly reduce significantly the memory and computational cost for low rates.

\subsection{Efficiency}
Lightweight architectures, such as GoogleNet~\cite{szegedy2015going} and MobileNet~\cite{howard2017mobilenets,sandler2018mobilenetv2}, are designed for resource-limited devices by reducing the number of parameters and computation. At the cost of small drop in performance, integer or binary weights can further improve efficiency~\cite{khan1994integer,rastegari2016xnor,jacob2018quantization}. Network architecture search (NAS)~\cite{zoph2016neural,baker2016designing,tan2019mnasnet} includes design hyperparameters (e.g. width, number of layers) in the optimization space. 
Slimmable neural networks~\cite{yu2019slimmable,yu2019universally} enable models that can be run at different accuracy-efficiency tradeoffs. Regarding NIC, Johnston \textit{et al.}~\cite{johnston2019computationally} use NAS to achieve 2-3$\times$ coding speed-up. Cai \textit{et al.}~\cite{cai2019novel} use progressive coding to reduce initial latency, although memory and computational cost remain similar. While tackling run-time or latency, these methods still focus on a single RD tradeoff, not providing rate, memory nor computation control.

\section{Slimmable compressive autoencoders}
\subsection{Slimmable autoencoders}
The basic structure of an autoencoder (AE) is a learnable encoder $\mathbf{z}=f\left(\mathbf{x};\theta\right)$ parametrized by $\theta$ that maps an input image $\mathbf{x}\in \mathbb{R}^{N}$ to a transformed (latent) representation $\mathbf{z}\in\mathbb{R}^{D}$, followed by a learnable decoder $\mathbf{\hat{x}}=g\left(\mathbf{z};\phi\right)$ parametrized by $\phi$ that maps the latent representation to $\mathbf{\hat{x}}\in \mathbb{R}^{N}$, with the objective of reconstructing the input image $\mathbf{x}$. Hence the combination of encoder and decoder is autoencoding $\mathbf{x}$, and the objective is to learn the parameters $\psi=\left(\theta,\phi\right)$ by minimizing a loss $\mathcal{L}\left(\theta,\phi;\mathcal{X}\right)$ given a training dataset $\mathcal{X}=\left\{\mathbf{x}_i\right\}_{i=1}^{|\mathcal{X}|}$. The loss measures the reconstruction error, possibly combined with other objectives.

We are interested in AEs whose layers are slimmable~\cite{yu2019slimmable}, i.e. \textit{slimmable autoencoders} (\textit{SlimAEs}), thus enabling dynamic control over the memory and computation costs. An \textit{slimmable layer} allows for discarding part of the layer parameters (in most cases is equivalent to setting them to zero) while still performing a valid operation, trading off expressiveness for lower memory and computational cost. We consider SlimAEs containing $K$ \textit{subautoencoders} (\textit{subAEs}), each of them parametrized by a pair $\psi^{(k)}=\left(\theta^{(k)},\phi^{(k)}\right)\in \Psi=\left\{\left(\theta^{(1)},\phi^{(1)}\right),\ldots,\left(\theta^{(K)},\phi^{(K)}\right)\right\}$, where we assume that $\theta^{(1)}\subset\!\cdots\!\subset\!\theta^{(K)}\!=\!\theta$ and $\phi^{(1)}\subset\cdots\subset\phi^{(K)}=\phi$ (we assume these conditions are met for every layer in the SlimAE). Similarly, we can define the loss for the subAE $k$ as $\mathcal{L}^{(k)}\left(\theta^{(k)},\phi^{(k)};\mathcal{X}\right)$, and train the SlimAE with the joint loss or a weighted average $\mathcal{L}\left(\Psi;\mathcal{X}\right)=\sum_{k}\mathcal{L}^{(k)}\left(\theta^{(k)},\phi^{(k)};\mathcal{X}\right)$.

\begin{figure}[ht]
	\centering
	\includegraphics[width=0.9\columnwidth]{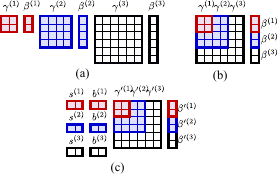}
	\vspace{-0.8em}
	\caption{GDN variants: (a) SwitchGDN, (b) SlimGDN, (c) SlimGDN+ (SlimGDN with switch. param. modulation).}
	\label{gdn_variants}
	\vspace{-1em}
\end{figure}

\subsection{Compressive autoencoders}
A compressive autoencoder (CAE) is an AE, where the output of the encoder is a binary stream (\textit{bitstream}), typically stored or transmitted through a communications channel. The objective is to maximize the quality of the reconstructed image (i.e. minimize the \textit{distortion}) while minimizing the number of bits transmitted (i.e. minimize the \textit{rate}). CAEs are based on AEs, where the encoder is followed by a quantizer $\mathbf{q}=Q\left(\mathbf{z}\right)$, where $\mathbf{q}\in \mathbb{Z}^D$ is a discrete-valued symbol vector. A losseless entropy encoder then binarizes and serializes $\mathbf{q}$ into the bitstream $\mathbf{b}$, exploiting its statistical redundancy to achieve code lengths close to its entropy. These operations are reversed in the decoder. 

CAEs are typically trained by solving a \textit{rate-distortion optimization (RDO)} problem with loss
\begin{equation}
    \mathcal{L}\left(\theta,\phi;\mathcal{X},\lambda\right)=D\left(\theta,\phi;\mathcal{X}\right)+\lambda R\left(\theta;\mathcal{X}\right),
    \label{eq:CAE_loss}
\end{equation}
where $\mathcal{X}$ is the training dataset, $\lambda$ is the (fixed) tradeoff  between rate and distortion. To allow end-to-end optimization with backpropagation, during training non-differentiable operations, such as quantization and entropy coding, are replaced by differentiable proxies, such as additive noise and entropy estimation.

Without loss of generality, we focus on the CAE framework of Balle \textit{et al.}~\cite{balle2016end}, which combines convolutional layers, generalized divisive normalization (GDN)  and inverse GDN (IGDN) layers, scalar  quantization to the nearest neighbor (i.e. i.e. $\mathbf{q}=\lfloor\mathbf{z}\rfloor$) and arithmetic coding. During training, quantization is replaced by additive uniform noise (i.e. $\mathbf{\tilde{z}}=\mathbf{z}+\Delta \mathbf{z}$, with $\Delta \mathbf{z}\sim \mathcal{U}\left(-\frac{1}{2},\frac{1}{2}\right)$). Similarly, arithmetic coding is bypassed and rate is approximated by the entropy of the quantized symbol vector $R\left(\mathbf{b}\right)\approx H\left[P_\mathbf{q}\right]\approx H\left[p_\mathbf{\tilde{z}}\left(\mathbf{\tilde{z}};\nu\right)\right]$, where $\nu$ are the parameters of the entropy model used in~\cite{balle2016end}. Distortion is measured as the reconstruction mean square error (MSE), i.e. $\left\|\mathbf{x}-\mathbf{\hat{x}}\right\|^2$. The CAE is thus parametrized by $\psi=\left(\theta,\phi,\nu\right)$.

\subsection{Slimmable CAEs}
In order to obtain a \textit{slimmable compressive autoencoder} (\textit{SlimCAE}),  all operations in the CAE are required to be non-parametric, slimmable or efficiently switchable. Quantization is non-parametric in our case, and convolutional layers are implemented slimmable~\cite{yu2019slimmable}. For GDN/IGDN layers, we propose and compare several variants (see next subsection) layers. Finally, we use switchable entropy models, i.e. each subCAE $k$ has its own parameters $\nu^{(k)}$, which can be easily switched since the size is negligible compared to the other parameters $\theta^{(k)}$ or $\phi^{(k)}$.

Our approach can also be extended to more complex frameworks including hyperpriors~\cite{balle2018variational} and autoregressive context models~\cite{minnen2018joint}.

\subsection{Switchable and slimmable GDN/IGDN layers}
While GDN~\cite{balle2016density} was proposed to Gaussianize the local joint statistics of natural images, Balle \textit{et al.}~\cite{balle2016end} proposed an approximate inverse operation (IGDN), and showed that GDN/IGDN layer pairs are highly beneficial in learned image compression, and since then have been adopted by many CAE frameworks. Both GDN and IGDN are parametrized by $\gamma\in \mathbb{R}^{w\times w}$ and $\beta\in\mathbb{R}^{w}$, where $w$ is the number of input (and output) channels.

In the case of a SlimCAE with $K$ subCAEs, the input to the GDN layer has the following possible channel dimensions $w^{\left(1 \right )},\ldots,w^{\left(K \right )}$. We consider three possible variants:
\begin{itemize}
    \item \textbf{Switchable GDNs\footnote{In the following, we omit IGDN for clarity (the same analysis applies).}}. We use independent sets of parameters $\gamma^{(k)}\in \mathbb{R}^{w^{(k)}\times w^{(k)}}$ and $\beta^{(k)}\in\mathbb{R}^{w^{(k)}}$ for every subGDN $k$ (see Fig.~\ref{gdn_variants} a). The normalized representation for an input $\mathbf{y}^{(k)}\in\mathbb{R}^{w^{(k)}}$ is then
    \begin{equation}
        \vspace{-0.3em}
        \tilde{y}_i^{(k)} = \frac{y_i^{(k)}}{\left(\beta_i^{(k)}+\sum_{j}\gamma_{ij}^{(k)}|y_j^{(k)}|^2\right)^{\frac{1}{2}}}
        \label{eq:GDNs}
        \vspace{-0.3em}
    \end{equation}
    While flexible, the total number of parameters is relatively high $\sum_{k=1}^K\left(w^{\left(k \right )}+1\right )w^{\left(k \right )}$, and switching may be not very efficient.  
    
    \item \textbf{Slimmable GDN (SlimGDN)}. A more compact option is to reuse parameters from smaller subGDNs by imposing $\gamma^{(1)}\subset\cdots\subset\gamma^{(K)}$ and $\beta^{(1)}\subset\cdots\subset\beta^{(K)}$. Now the total number of parameters in a SlimGDN layer is $\left(M^{\left(K \right )}+1\right )w^{\left(K \right )}$  (see Fig.~\ref{gdn_variants}b).
    
    \item \textbf{SlimGDN with switchable parameter modulation}. SlimGDNs usually performs worse than switchable GDNs, since they are less flexible to adapt to the statistics of the different $\mathbf{y}^{(k)}$. We propose a variant using switchable parameter modulation, where a global scale and bias are learned separately for every subGDN (i.e. switchable), i.e. $\gamma_{ij}^{(k)}=s_\gamma^{(k)}{\gamma'}_{ij}^{(k)}+b_\gamma^{(k)}$ and $\beta_i^{(k)}=s_\beta^{(k)}{\beta'}_i^{(k)}+b_\beta^{(k)}$, where ${\gamma'}^{(k)}$ and ${\beta'}^{(k)}$ are shared and slimmable and $s_\gamma^{(k)}$, $b_\gamma^{(k)}$, $s_\beta^{(k)}$ and $b_\beta^{(k)}$ are switchable scalars specific for the subGDN $k$. This variant requires only 4 additional parameters per subGDN  (see Fig.~\ref{gdn_variants}c), for a total number of parameters $\left(w^{\left(K \right )}+1\right )w^{\left(K \right )}+4K$.
\end{itemize}

\subsection{(Naive) training of SlimCAEs}\label{sec:naive_training}
We can extend Eq.~(\ref{eq:CAE_loss}) and optimize the joint loss of all $K$ subCAEs $\arg \min_\psi \sum_{\psi \in \Psi}\mathcal{L}\left(\psi;\mathcal{X},\lambda\right)$, with parameters
$\Psi=\left\{\psi^{(1)},\ldots,\psi^{(K)}\right\}$. The problem can be solved using stochastic gradient descent (SGD) and backpropagation. We refer to this case as \textit{naive SlimCAE}.

\section{SlimCAEs with multiple rate-distortion tradeoffs}
\subsection{Rate-distortion and capacity}
\begin{figure}[ht]
	\centering
	\includegraphics[width=\columnwidth]{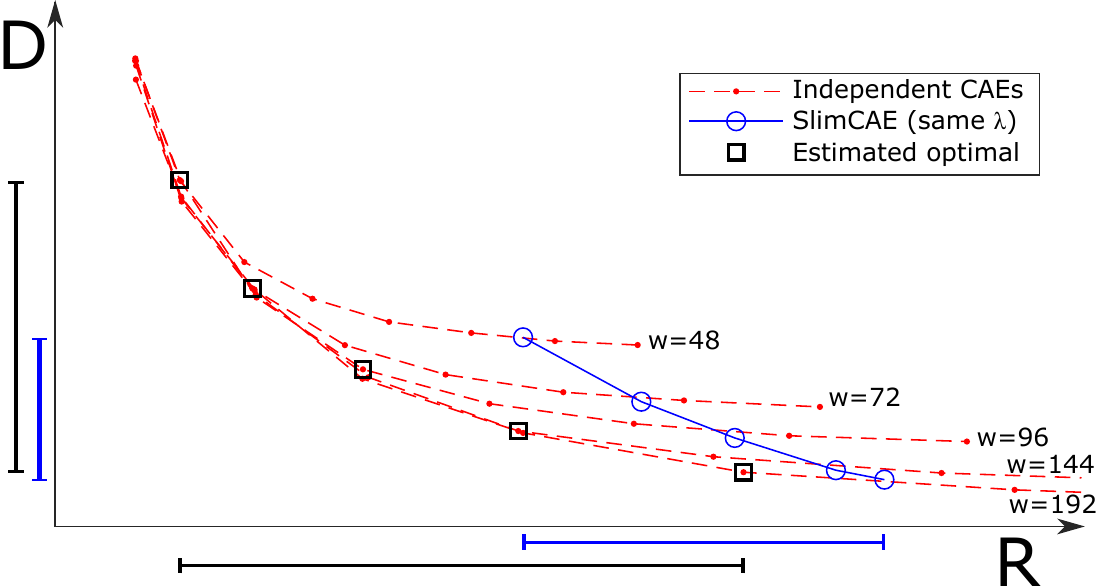}
	\vspace{-0.8em}
	\caption{Comparison of RD curves of independent CAEs and SlimCAE with shared $\lambda$. Better choices in the RD curves are also highlighted.}
	\vspace{-1em}
	\label{fig:naive_RDcurve}
\end{figure}

While a naive SlimCAE can already control the rate of the output bitstream and the complexity of the model, it is limited to a relatively narrow range of rates with suboptimal RD performance. This can be observed in Fig.~\ref{fig:naive_RDcurve}, where we show the RD curves obtained with independent CAEs with different capacities controlled via the layer width $w$ (i.e. number of filters per convolutional layer) compared with one SlimCAE with the same layer widths.

Fig.~\ref{fig:naive_RDcurve} shows that there is a limit in the minimum achievable distortion by a CAE (i.e. at high rates), which is in turn related to its capacity (the lower the capacity, the higher the minimum distortion). Additionally, the figure shows that, when the rate is low enough, additional capacity is unnecessary since the curves converge before that point, (i.e. an optimal capacity for every segment of the optimal RD curve).

Note also that the RD points of the SlimCAE are located over the RD curves of independent CAEs with the same capacity, suggesting that a slimmable version does not entail RD penalty. We aim at training the SlimCAE so it can achieve optimal RD performance at the different capacities. We define the \textit{multi-RD optimization} (MRDO) loss as 
\begin{equation}
    \vspace{-0.3em}
    \mathcal{L}\left(\Psi;\mathcal{X},\Lambda\right)=\sum_{k=1}^K D\left(\psi^{(k)};\mathcal{X}\right)+\lambda^{(k)} R\left(\psi^{(k)};\mathcal{X}\right),
    \label{eq:SlimCAE_multiRDO_loss}
    \vspace{-0.3em}
\end{equation}
where $\Lambda=\left\{\lambda^{(1)},\ldots,\lambda^{(K)}\right\}$ is a given set of RD tradeoffs for the different $K$ subCAEs, and the corresponding MRDO problem is ${\arg \min}_\Psi \sum_{\psi^{(k)} \in \Psi}\mathcal{L}\left(\Psi;\mathcal{X},\Lambda\right)$. 

An important aspect to note is that in this case each subCAE solves a different optimization problem determined by the specific tradeoff $\lambda^{(k)}$. This is an important difference with slimmable networks and SlimAEs in general (including naive SlimCAEs), where every subnetwork or subAE solves exactly the same problem, just with different capacity. This has implications, such as more difficulty to jointly solve all the subproblems and also makes implicit distillation~\cite{yu2019slimmable} across subCAEs more unlikely. The problem now is to find appropriate values of $\Lambda$ and train the SlimCAE. 

\subsection{Estimating optimal $\lambda$s from RD curves}\label{sec:estimated_lambda}
One possible way is to leverage the RD curves of independent CAEs and try to estimate the optimal points to switch to the next subCAE, which is where the curves start diverging because RD performance saturates for that capacity. We can then estimate $\lambda^{(k)}$ as the slope of the corresponding curve at that optimal point (see Fig.~\ref{fig:naive_RDcurve}).

\subsection{Automatic estimation via $\lambda$-scheduling}
\begin{figure}[t]
	\centering
	\includegraphics[width=\columnwidth]{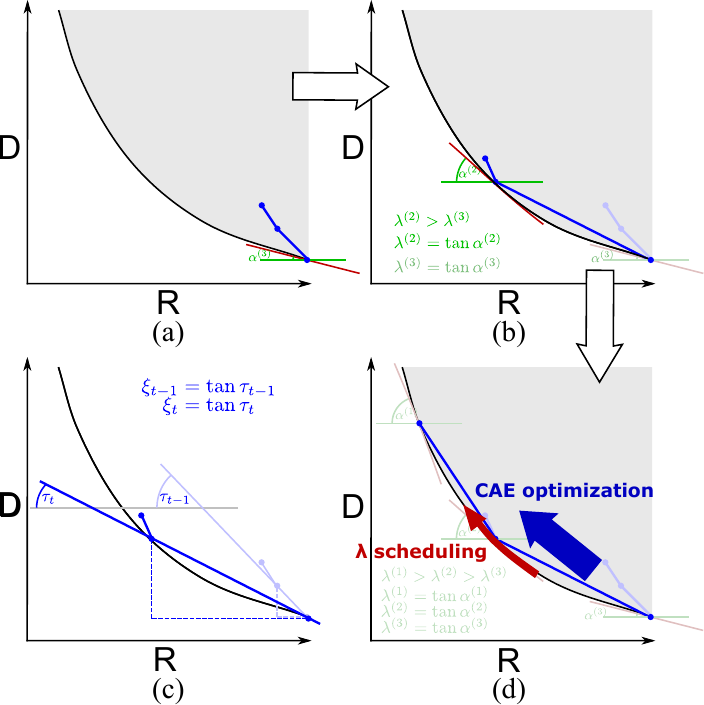}
	\vspace{-0.8em}
	\caption{Training SlimCAEs: (a) naive training with a single RD tradeoff $\lambda$ leads to small ranges and suboptimal RD performance, (b) varying the $\lambda$ progressively for smaller subCAEs changes the target RD point and stretches the RD range, (c) the slope $\xi$ of RD segments is used to monitor convergence of the proposed training algorithm, and (d) illustration of how $\lambda$ scheduling changes the RD target and SlimCAE optimization stretches the RD range.}
	\label{fig:training_algorithm}
	\vspace{-1.2em}
\end{figure}

While knowing in advance the empirical RD curves leads to better RD performance and wider rate ranges, it has the important drawback of having very high computation cost, since we need to compute $K$ RD curves, one for each target capacity, and every curve requires training a number of independent CAE exploring different $\lambda$s.

\textit{MRDO training with $\lambda$ scheduling}\label{sec:lambda_scheduling}
In order to address the previous limitation, we propose an effective MRDO training algorithm to automatically estimate $\Lambda$ without requiring independent CAEs curves (see Alg.~\ref{alg:training_with_scheduling}).

The algorithm is based on progressively varying the values of every $\lambda^{(k)}$ following a predefined schedule. We alternate phases of updating $\Lambda$ and updating the SlimCAE using SGD. The initial stage is a naive SlimCAE, where $\lambda$ is set to target high rate and low distortion, which requires full use of the capacity. Once trained, the SlimCAE is already optimized for that full capacity. We fix $\lambda^{(K)}$ and update the remaining $k=1,\ldots,K-1$ as $\lambda_{t}^{(k)}=\kappa\lambda_{t-1}^{(k)}$ with a factor $\kappa>1$. Then we update the SlimCAE for another number of iterations, which tends to reduce the rate and moves the RD of subCAE $K-1$ closer to the optimal RD curve. Geometrically, this results in the slope of the segment between the RD points of the two consecutive subCAEs $K-1$ and $K$ decreasing (see Fig.~\ref{fig:training_algorithm}b and c). When this slope does not decrease anymore, we fix $\lambda^{(K-1)}$ and continue the process recursively. The overall effect of the $\lambda$ scheduling is to progressively accommodate the target RD point for each subCAE so training can approximate the optimal RD points. 

\begin{algorithm}
\caption{SlimCAE training with $\lambda$-scheduling}\label{alg:training_with_scheduling}
\renewcommand{\algorithmicrequire}{\textbf{Input:}}
\renewcommand{\algorithmicensure}{\textbf{Output:}}
\begin{algorithmic}[1]
	\Require $\mathcal{X}_{tr}$, $\mathcal{X}_{val}$ \Comment{Training, val. data}
	\Require SlimCAE with $K$ subCAEs and parameters $\Psi$
	\Require $\lambda^{(K)}$ \Comment{RD tradeoff for largest subCAE}
	\Require $\kappa$, $T$, $M$ \Comment{Other hyperparameters}
	\Ensure $\Psi,\Lambda$
	\State $\Lambda_0 \gets \left[\lambda^{(K)},\ldots,\lambda^{(K)}\right]$
	\State $\Psi_0 \gets {\arg\min}_\Psi\mathcal{L}\left(\Psi;\mathcal{X}_{tr},\Lambda_0\right)$  \Comment{Naive training}
    \State Calculate $R^{(K-1)}_0\!,R^{(K)}_0\!,\!D^{(K-1)}_0,\!D^{(K)}_0$ over $\mathcal{X}_{val}$
    \State $\xi_0 \gets \frac{D_0^{(K)}-D_0^{(K-1)}}{R_0^{(K)}-R_0^{(K-1)}}$
    \State $t\gets 1$
    \For{$i\gets K-1$ to $1$}
        \For {$m\gets 1$ to $M$}
            \State $\Lambda_i\gets \left[
            \kappa\lambda_{i-1}^{(1)},\ldots,\kappa\lambda_{i-1}^{(i)},\lambda_{i-1}^{(i+1)},\ldots,\lambda_{i-1}^{(K)}\right]$
            \For {$j\gets 1$ to $T$}
                \State $\Psi_t \gets {\arg\min}_\Psi\mathcal{L}\left(\Psi_{t-1};\mathcal{X}_{tr},\Lambda_i\right)$ 
                \State $t\gets t+1$
            \EndFor
            \If {$R_t^{(i+1)} < R_t^{(i)}$}
                \State \textbf{continue}
            \Else{}
                \State Calculate $\xi_t \gets  \frac{D_t^{(i+1)}-D_t^{(i)}}{R_t^{(i+1)}-R_t^{(i)}}$ over $\mathcal{X}_{val}$
                \If{$\xi_t>\xi_{t-T}$}
                    \State \textbf{break}
                \EndIf
            \EndIf
        \EndFor
    \EndFor

	\end{algorithmic}
\end{algorithm}

\section{Experiments}
\subsection{Experimental settings}
We implemented\footnote{\url{https://github.com/FireFYF/SlimCAE}} and evaluated the proposed approaches building upon the widely used image compression framework proposed by Balle at al.~\cite{balle2016end} which typically uses layers with a width of 192 filters (encoder: 3 conv, 3 GDN; decoder: 3 deconv, 3 IGDN). To address different complexities and rates, we consider five different widths ($w\in \left\{48,72,96,144,192\right\}$), which also control the total capacity of the model. The models are trained and evaluated on the CLIC dataset\footnote{\url{https://www.compression.cc/2019/challenge}}. In  Section~\ref{ssec:slimmable_entropy} we evaluate on the Kodak
\footnote{\url{http://r0k.us/graphics/kodak}}
and Tecnick\footnote{\url{https://testimages.org/sampling}} datasets~ to compare to other methods. We evaluate both RD performance and efficiency (memory footprint, computational cost and latency).

\textbf{SlimCAE variants.} We consider three GDN/IGDN variants (SwitchGDN, SlimGDN and SlimGDN+ corresponding to Fig.~\ref{gdn_variants}a, b and c respectively) and three training strategies (naive, estimated $\lambda$s and $\lambda$-scheduling, corresponding to the methods described in Sections~\ref{sec:naive_training}, \ref{sec:estimated_lambda} and \ref{sec:lambda_scheduling}). We used a switchable entropy model (i.e. one independent entropy model per width), since the number of parameters is negligible compared to the overall model.

\textbf{Baselines.} We compare SlimCAE to independently trained CAEs for different widths (five models in total following \cite{balle2016end}). We also compare to three approaches to provide variable rate in a single model: bottleneck scaling (BScale)~\cite{theis2017lossy}\footnote{Note that \cite{theis2017lossy} actually introduces the term \textit{compressive autoencoder}. s in a general sense, and in our experiments CAE refers to our baseline~\cite{balle2016end}, while BScale denotes the variable rate approach in \cite{theis2017lossy}}, modulated autoencoders (MAE)~\cite{yang2020variable} and conditonal autoencoders (cAE)~\cite{choi2019variable}.

\textbf{Training details.} We use $240\times240$ pixel crops and a batch size of 8. Some methods are trained in one step, while other require two steps. The former includes independent CAEs, MAE, cAE, SlimCAE with naive training and training with estimated $\lambda$s. In these cases we use a learning rate of 1e-4 (1e-3 for the entropy model) during 1.2M iterations, and then halve them for an additional 200K iterations. SlimCAE with $\lambda$-scheduling uses a SlimCAE after naive training 1.2M iteration, followed by training with $\lambda$-scheduling ($\kappa=0.8$, $T=2000$ and $M=7$ in  Alg.~\ref{alg:training_with_scheduling})\footnote{We extend the implementation of~\cite{balle2016end}, which optimizes $\lambda D + R$. We adapt $\lambda$-scheduling correspondingly.} during 28K iterations and then fine tuned (by halving the learning rates) until a total of 1.5M iterations with the final $\lambda$s fixed. We measure distortion as MSE during training and as PSNR during $\lambda$-scheduling. BScale uses a CAE trained during 1.2M iterations, which is then fixed and scaling parameters are learned during 300K iterations.

\subsection{Qualitative analysis}
\begin{figure}[t]
    \centering
	\includegraphics[width=\columnwidth]{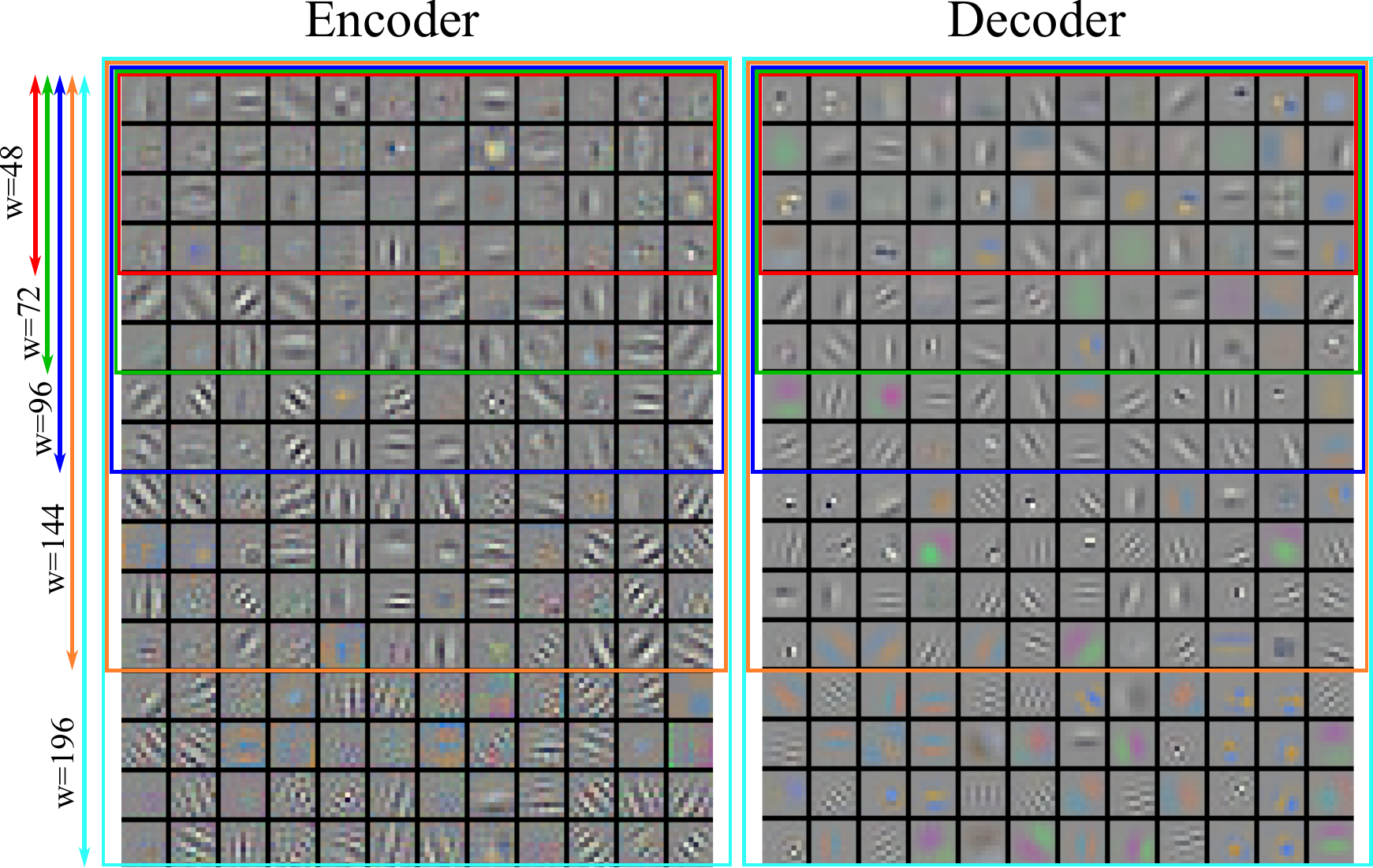}
	\vspace{-0.8em}
    \caption{Filters in the first convolutional layer (encoder) and last convolutional layer (decoder) for different widths.}
	\label{fig:filters}
	\vspace{-1.4em}
\end{figure}

SlimCAE can effectively distribute and optimize the capacity in a way that each subCAE can focus on the patterns relevant to its own optimal RD tradeoff. For example, the first convolutional layer of the smallest subencoder (see Fig.~\ref{fig:filters}) contains only filters sensitive to low frequency patterns, while larger subencoders progressively include filters related with higher frequency, since they are necessary to achieve lower distortion. The latent representation in the bottleneck is also structured in a similar way from coarse reconstruction to fine details (see Fig.~\ref{latent} a-b in supp. mat.).

\subsection{Rate-distortion}
\begin{figure*}[t]
\vspace{-0.8em}
    \centering
    \begin{subfigure}[b]{0.33\textwidth}
         \centering
         \includegraphics[width=\textwidth]{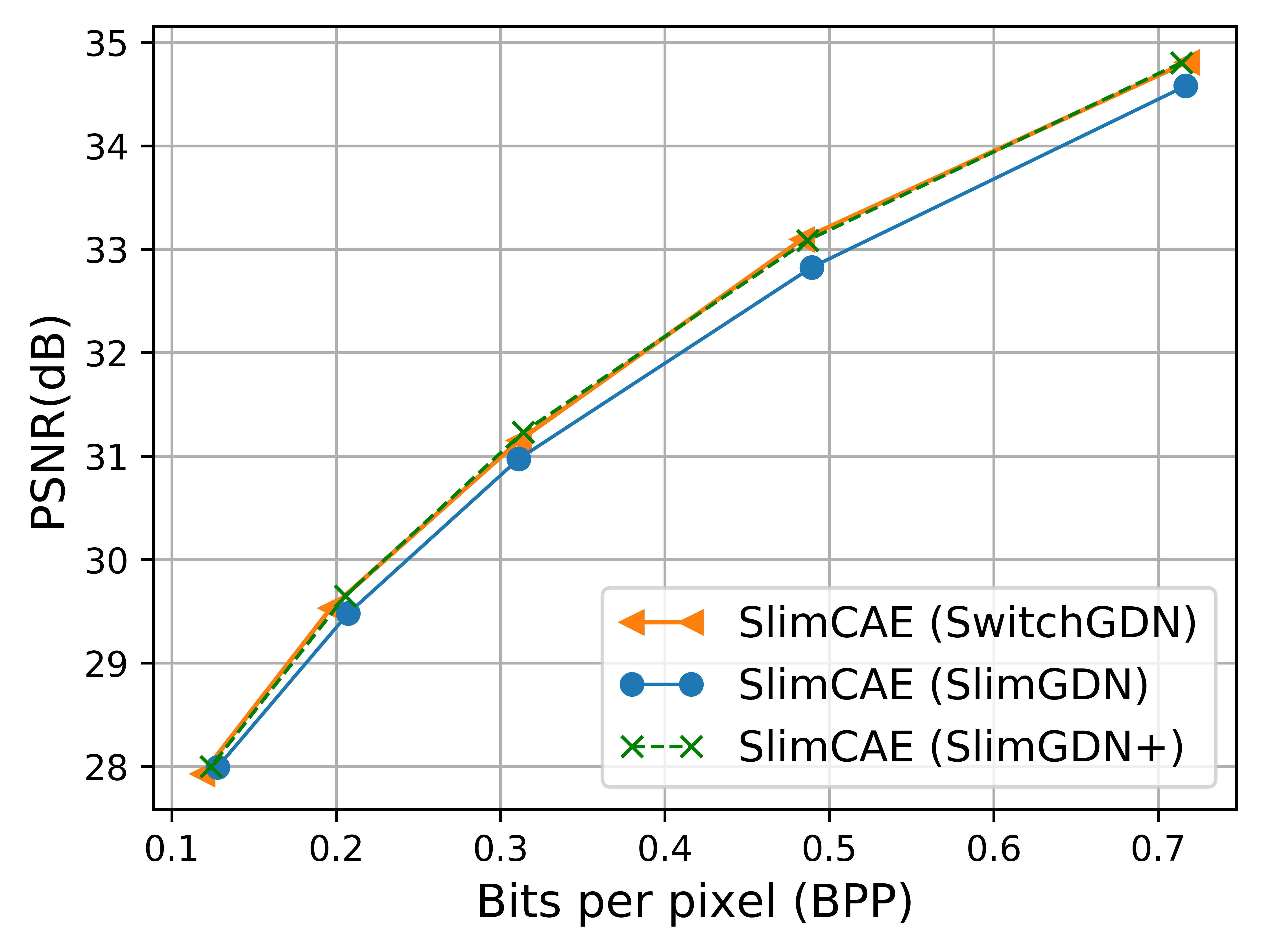}
         \caption{GDN variants (with estimated $\lambda$s)}
         \label{sfig:rd_GDNs}
     \end{subfigure}
     \hfill
         \begin{subfigure}[b]{0.33\textwidth}
         \centering
         \includegraphics[width=\textwidth]{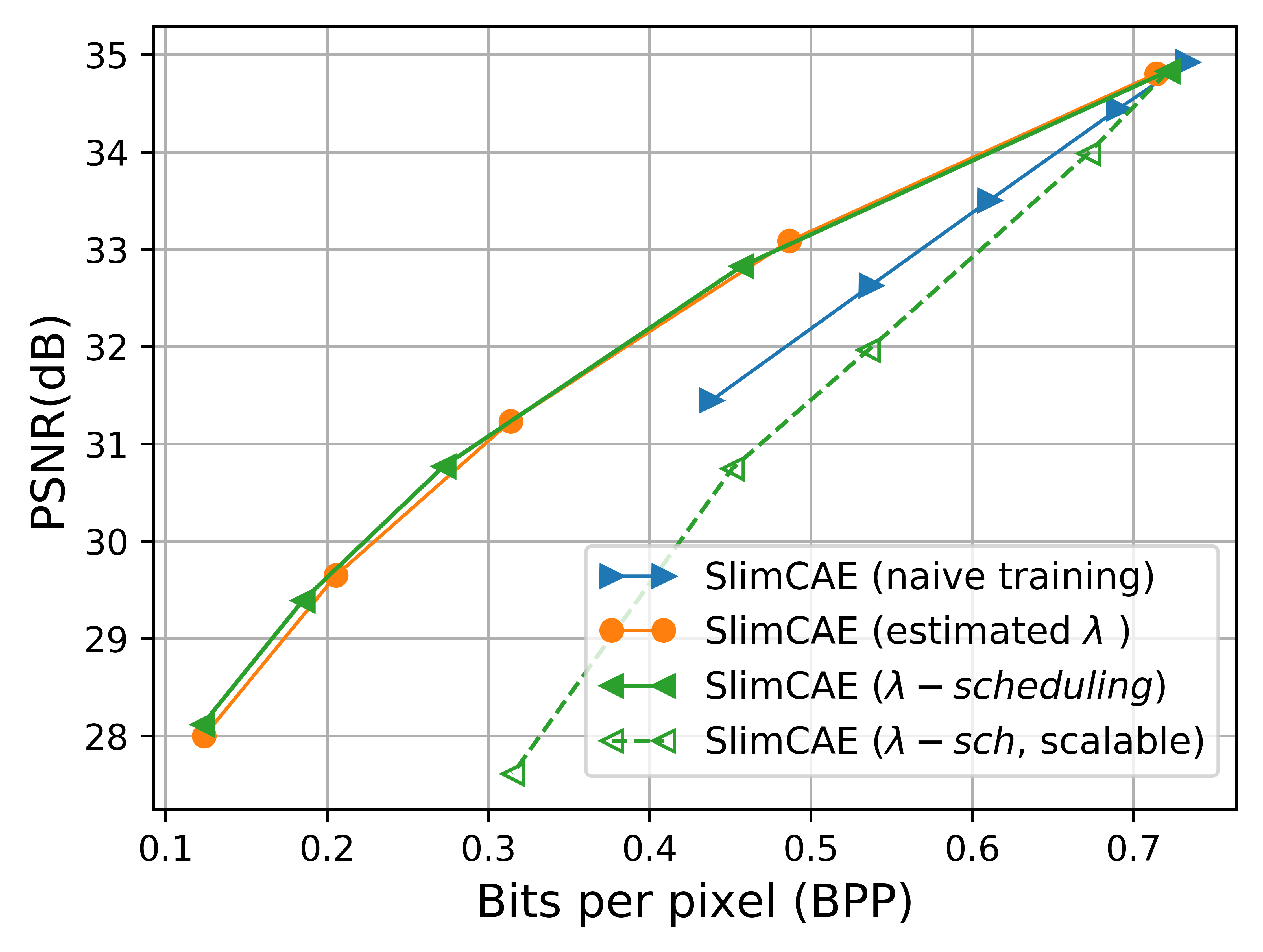}
         \caption{Training strategies and scalability}
         \label{sfig:rd_training_strategies}
     \end{subfigure}
     \hfill
     \begin{subfigure}[b]{0.33\textwidth}
         \centering
         \includegraphics[width=\textwidth]{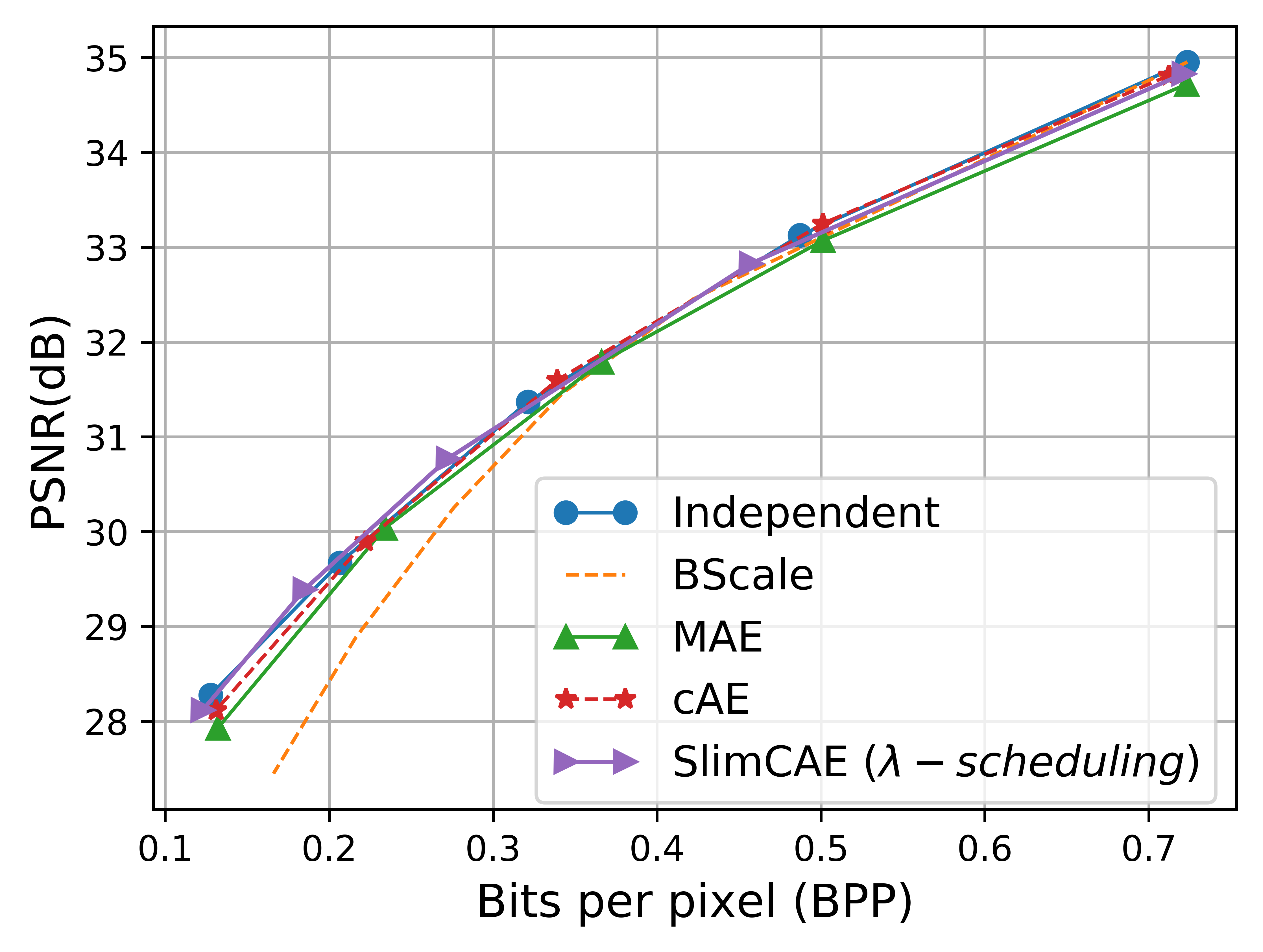}
         \caption{Variable rate methods}
         \label{sfig:rd_variable_rate_baselines}
     \end{subfigure}
     \vspace{-1.7em}
	\caption{Rate-distortion performance comparison (CLIC dataset).}
	\label{fig:rd_SlimCAE}
	\vspace{-0.8em}
\end{figure*}
Fig.~\ref{sfig:rd_GDNs} shows the rate-distortion performance obtained with different GDN variants. Sharing GDN parameters across different widths (i.e. SlimGDN) results in worse performance than independent ones (i.e. SwitchGDN). However, this loss can be recovered when parameter modulation (i.e. SlimGDN+) at a negligible parameter increase.

Fig.~\ref{sfig:rd_training_strategies} shows that naive training suffers from the limitations of using a single shared tradeoff $\lambda$, while training with more adequate width-specific $\lambda$s (those estimated in Fig.~\ref{fig:naive_RDcurve}) results in an RD curve closer to the obtained with independent CAEs. Fig.~\ref{sfig:rd_training_strategies} also illustrates the effect of $\lambda$-scheduling in the RD curve, which gets progressively closer until it essentially achieves the same performance as independent CAEs (see Fig.~\ref{sfig:lambda-sched} in supplementary for more details), but without requiring training auxiliary models.

Finally, we compare SlimCAE to other baselines enabling variable rate in a single model. SlimCAE obtains the best RD performance, overlapping with that of independent CAEs, but with a much lower training and memory cost, as we see next.

\subsection{Efficiency}
We also evaluate the efficiency of SlimCAE in terms of memory footprint (in MB), computational cost (in FLOPs) and latency (in ms)\footnote{We only compare to NIC codecs, since traditional codecs run in different hardware (CPU instead of GPU). As reference, our BPG baseline takes on CPU (for 0.1-1.0 bpp) 3.2-4.5 s/img (enc) and 95-131 ms/img (dec).}. Values in features and parameters are represented with 4 bytes, and the features are calculated for input images of size $768\times512$ pixels. We consider a baseline with five independent CAEs optimized for different RD tradeoffs. For fair comparison we use the minimum width that guarantees that the RD performance at a particular tradeoff $\lambda$ remains optimal (see Fig.~\ref{fig:naive_RDcurve}).

\begin{table}[!ht]
\centering
\vspace{-0.8em}
  \caption{Computational cost of trained models (millions of FLOPs). Some methods adjust layer widths.}
  \label{FLOPS}
  \resizebox{83mm}{15mm}{
  \begin{tabular}{l|ccccc}
    \hline
    Methods       &Low rate & $\rightarrow$ & Medium rate & $\rightarrow$ &High rate  \\
    \hline
    \multirow{2}{*}{Independent} &15.34\textbf{M} &31.69\textbf{M} &53.81\textbf{M} &115.53\textbf{M} &200.28\textbf{M} \\
                                 &(w=48)         &(w=72)          &(w=96)          &(w=144)          &(w=192) \\
    \multirow{2}{*}{BScale~\cite{theis2017lossy}} &200.28\textbf{M} &200.28\textbf{M} &200.28\textbf{M} &200.28\textbf{M} &200.28\textbf{M} \\
                                 &(w=192)       &(w=192)    &(w=192)     &(w=192)    &(w=192) \\
    \multirow{2}{*}{MAE~\cite{yang2020variable}} &200.40\textbf{M} &200.40\textbf{M} &200.40\textbf{M} &200.40\textbf{M} &200.40\textbf{M} \\
                                 &(w=192)       &(w=192)    &(w=192)     &(w=192)    &(w=192) \\
    \multirow{2}{*}{cAE~\cite{choi2019variable}} &200.31\textbf{M} &200.31\textbf{M} &200.31\textbf{M} &200.31\textbf{M} &200.31\textbf{M}\\
                                 &(w=192)       &(w=192)    &(w=192)     &(w=192)    &(w=192) \\        
    \hline
    \multirow{2}{*}{SlimCAE} &\textbf{15.34M} &\textbf{31.69M} &\textbf{53.81M} &\textbf{115.53M} &\textbf{200.28M}\\
            &(w=48)       &(w=72)    &(w=96)     &(w=144)    &(w=192) \\
    \hline
    \end{tabular}}
    \vspace{-0.2em}
\end{table}

\begin{table}[!ht]
\centering
\vspace{-0.8em}
  \caption{Encoding and decoding latency (ms) for a $768\!\times\!512$ input image (i.e. batch size 1) on a NVIDIA GTX 1080Ti GPU (excluding data loading/writing and arithmetic coding.}
  \label{latency}
  \resizebox{83mm}{16mm}{
  \begin{tabular}{ll|ccccc}
    \hline
    \multicolumn{2}{c|}{Methods}   &Low rate & $\rightarrow$ & Medium rate & $\rightarrow$ &High rate  \\
    \hline
    \multirow{5}{*}{\rotatebox[origin=c]{90}{Encoding}} &Independent & $1.9\pm0.19$ & $2.2\pm0.17$ &$2.8\pm0.22$ &$4.0\pm0.19$ &$5.1\pm0.20$ \\
    & BScale~\cite{theis2017lossy} &$5.2\pm0.11$ &$5.2\pm0.16$ &$5.2\pm0.22$ &$5.2\pm0.15$ &$5.2\pm0.13$ \\
    & MAE~\cite{yang2020variable} &$5.4\pm0.20$ &$5.4\pm0.20$ &$5.4\pm0.13$ &$5.4\pm0.10$ &$5.4\pm0.11$ \\
    & cAE~\cite{choi2019variable} &$5.5\pm0.18$ &$5.5\pm0.11$ &$5.5\pm0.14$ &$5.5\pm0.21$ &$5.5\pm0.28$\\
    & SlimCAE &$\textbf{1.9 $\pm$ 0.15}$ &$\textbf{2.2 $\pm$ 0.27}$ &$\textbf{2.8 $\pm$ 0.12}$ &$\textbf{4.0 $\pm$ 0.17}$ &$\textbf{5.1 $\pm$ 0.10}$\\
    \hline
    \multirow{5}{*}{\rotatebox[origin=c]{90}{Decoding}} &Independent &$2.9\pm0.20$ &$3.5\pm0.10$ &$4.3\pm0.07$ &$6.1\pm0.07$ &$8.0\pm0.13$ \\
    & BScale~\cite{theis2017lossy} &$8.0\pm0.21$ &$8.0\pm0.13$ &$8.0\pm0.18$ &$8.0\pm0.11$ &$8.0\pm0.13$ \\
    & MAE~\cite{yang2020variable} &$8.4\pm0.15$ &$8.4\pm0.13$ &$8.4\pm0.08$ &$8.4\pm0.07$ &$8.4\pm0.14$ \\
    & cAE~\cite{choi2019variable} &$8.5\pm0.20$ &$8.5\pm0.07$ &$8.5\pm0.09$ &$8.5\pm0.13$ &$8.5\pm0.21$ \\
    & SlimCAE &$\textbf{2.9 $\pm$ 0.10}$ &$\textbf{3.5 $\pm$ 0.07}$ &$\textbf{4.3 $\pm$ 0.10}$ &$\textbf{6.1 $\pm$ 0.16}$ &$\textbf{8.0 $\pm$ 0.13}$\\
    \hline
    \end{tabular}}
    \vspace{-0.8em}
\end{table}
\begin{figure}[ht]
  \centering
  \includegraphics[width=\columnwidth]{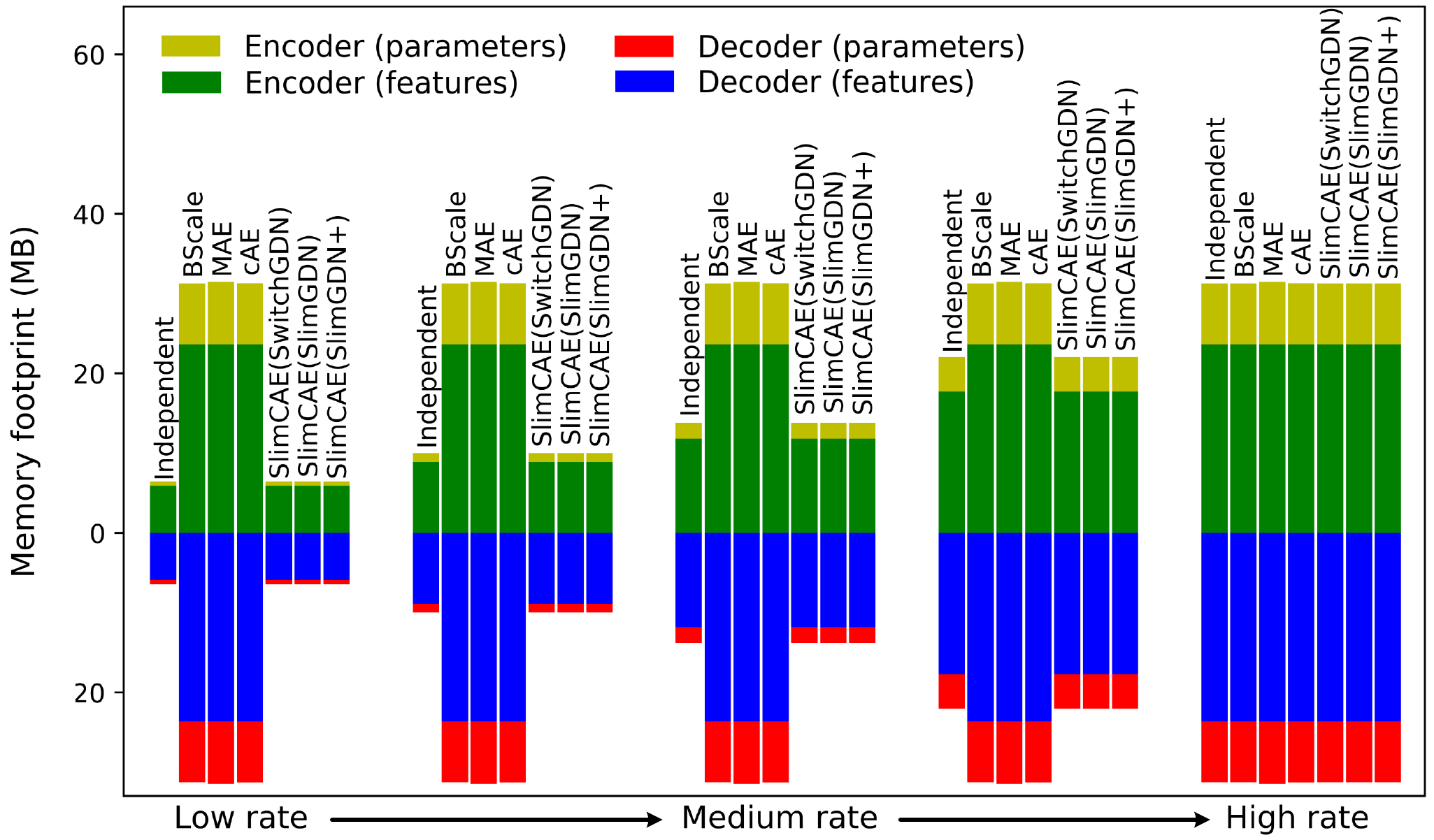}
  \caption{Memory footprint comparison for different rates.}
  \label{sfig:memory_footprint}
  \vspace{-0.8em}
\end{figure}
Fig.~\ref{sfig:memory_footprint} shows the memory footprint in the encoder and decoder at different widths. Features require significantly more memory than model parameters, especially at small widths. While for the largest width all methods require similar memory, an independent CAE and SlimCAE can reduce significantly the memory footprint for small widths. This reduction also results in a significantly lower computational cost (see Table~\ref{FLOPS}), and much lower latency during both encoding and decoding (see Table~\ref{latency}).  Now we consider the total memory required to provide the five different rates. It requires to store the model parameters of the independent CAEs (31.1 MB), in contrast to just a single model for BScale (15.3 MB), MAE (15.7 MB), cAE (15.3 MB) and SlimCAE (15.3 MB with SlimGDN+). If we consider the memory used to store features (note that at only one model is use at a time), the memory footprint of multiple CAEs varies from 42.9 to 78.3 MB, depending on the selected rate, and similarly to SlimCAE (27.1 to 62.5 MB). In contrast, other methods cannot adapt the complexity and remain with a higher and constant footprint (BScale 62.5 MB, MAE 63 MB and cAE 62.6 MB)\footnote{Note that, in practice, an optimized implementation could save some memory by discarding intermediate features once they are processed.}. The SlimGDN+ layers require 0.85 MB (compared to 1.71 MB and 0.85 MB in SwitchGDN and SlimGDN, respectively). 

Finally, Table~\ref{tab:comparison_methods} summarizes the main advantages and drawbacks of different methods. SlimCAE is the most complete of them providing variable rate with a single model and controllable memory and computational requirements, while achieving optimal RD performance. Training and switching between multiple CAEs suffers from a higher memory footprint (that increases with the number of target RD points), and the much higher cost of training multiple models. The other baselines can adapt rate, but not memory and computational costs, which remain high at low rates.

\subsection{Scalable bitstreams}
Motivated by SlimCAE's structured latent respresentation, we consider a variant where each group of channels are encoded independently, allowing \textit{quality scalability} and progressive decoding. The resulting bitstreams (i.e. base[+ enhancement stream(s)]) are all decodable by the SlimCAE decoder. On the other hand, the SlimCAE is no longer slimmable, so enabling quality scalability disables memory and computation scalability since the SlimCAE is no longer slimmable, and also has certain penalty in RD performance (see Fig.~\ref{sfig:rd_training_strategies} and Fig.~\ref{latent} c in supp. mat), a usual compromise in scalable image and video coding~\cite{taubman2000high,radha2001mpeg,schwarz2007overview}.

\subsection{Slimmable entropy models}\label{ssec:slimmable_entropy}
\begin{figure}
\vspace{-0.8em}
\centering
\includegraphics[width=\columnwidth]{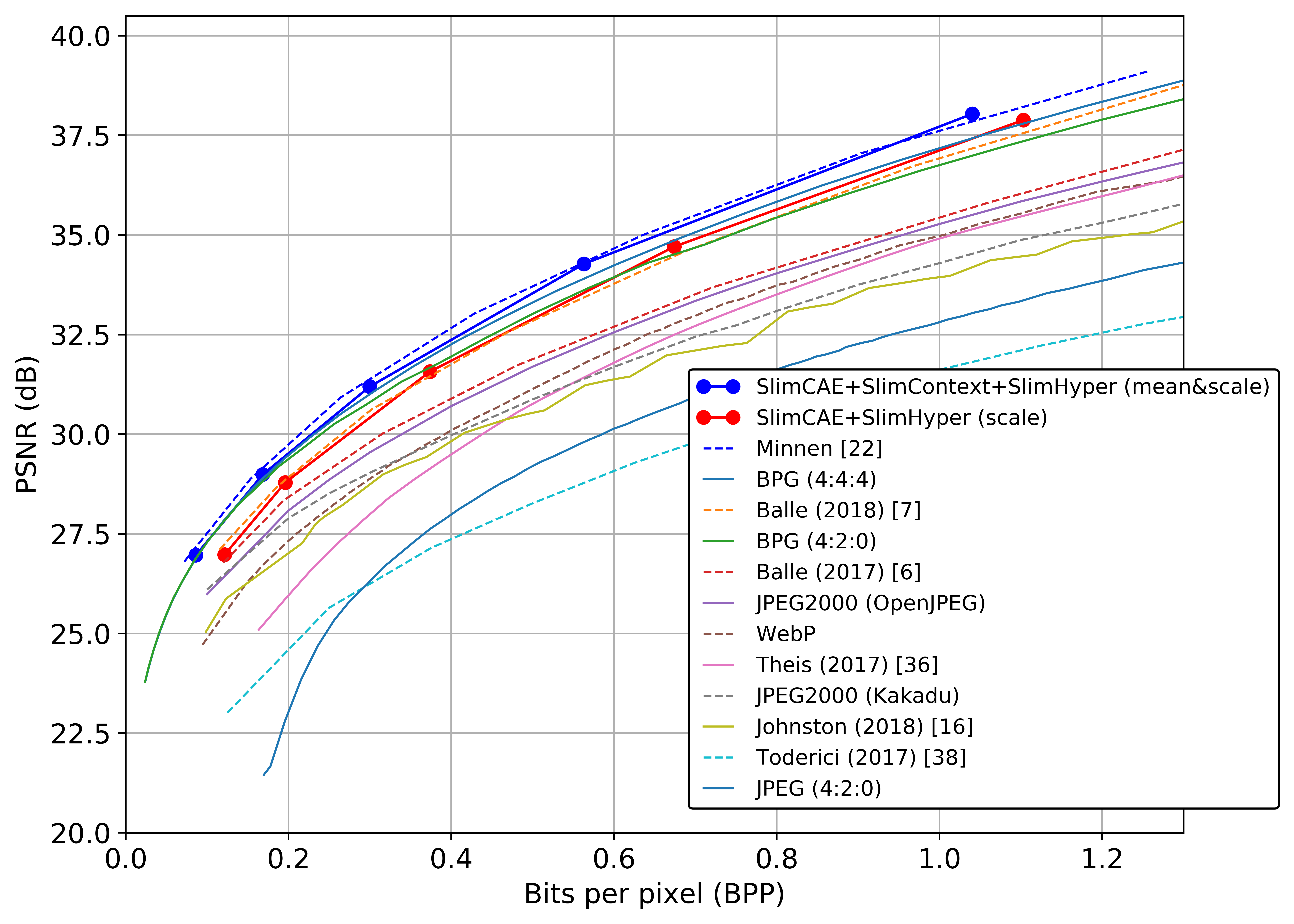}
\caption{Rate-distortion performance of SlimCAE with slimmable entropy model (Kodak dataset).
\textbf{\vspace{-0.8em}}
\label{fig:sota}}
\end{figure}

\begin{table}
\centering
\setlength{\tabcolsep}{2pt}
  \caption{BD-rate (\%) over BPG. 
  Lower means better.}
  \label{tab:BD-rate_adaptation}
  \vspace{-.2em}
  \resizebox{\columnwidth}{!}{
  \begin{tabular}{c|cc|cc|cc|cc}
    \hline
    \multirow{2}{*}{Dataset}&\multicolumn{4}{c|}{PSNR (opt. for MSE)} &\multicolumn{4}{|c}{MS-SSIM (opt. for MS-SSIM)} \\
    &[7] &[23] &Slim[7] &Slim[23] &[7] &[23] &Slim[7] &Slim[23] \\
    \hline
    Kodak &9.68 &-8.94 &9.52 &-6.17 &-41.46 &-46.92 &-41.41 &-47.88\\
    Tecnick &2.95 &-11.77 &5.23 &-10.43 &-41.50 & -43.24 &-40.34 &-48.94 \\
    \hline
    \end{tabular}
    }
    \vspace{-.6em}
\end{table}
	
Our approach is general and can be easily extended including slimmable versions of entropy models to achieve state-of-the-art RD performance. Following \cite{balle2018variational,minnen2018joint}, we train\footnote{~3M iter., batch 8, $256\times256$ crops, $\lambda$-sched ($\kappa\!=\!0.8$,$T\!=\!10^4$,$M\!=\!7$).} a larger capacity autoencoder\footnote{$w\in\left\{96, 144,192,288,384\right\}$} with a three conv layer slimmable  hyperprior\footnote{$w\in\left\{48,72,96,144,192\right\}$, and Leaky ReLUs (same in decoder)}~\cite{balle2018variational} and conditional convolutions~\cite{choi2019variable}. We also include a slimmable autoregressive context model\footnote{One masked conv layer with $w\in\left\{96,144,192,288,384\right\}$}~\cite{minnen2018joint}. We trained these models\footnote{On CLIC extended with 20k high quality images from \url{flickr.com}} and evaluated on Kodak and Tecnik datasets. Fig.~\ref{fig:sota} shows that our approach can be integrated with almost no penalty in RD performance, while providing the aforementioned advantages in terms of rate and complexity control of SlimCAEs. The same conclusions hold when optimizing MS-SSIM, keeping a significant gain over BPG (see Table~\ref{tab:BD-rate_adaptation}).

\section{Conclusion}
	Neural image compression is a new paradigm for image (and by extension video) coding, with numerous advantages over the traditional handcrafted linear transform coding. However, current approaches are also resource demanding, and usually tied to a particular rate, which limits their application in practice.

	Our approach is thus a step further towards practical and adaptive learned image compression, combining in a single model important functionalities, such as excellent rate-distortion performance, low and dynamically adjustable memory footprint, computational cost and latency, all of them easily controlled via a lightweight switching mechanism. This makes our approach attractive to resource-limited devices (e.g. smartphones), when rate and computation needs to be controlled dynamically (e.g. video coding, multi-tasking) or to deploy different models to heterogeneous devices, adapted to their computational capabilities.
	SlimCAE can also generate scalable bitstreams, which can be useful in streaming and broadcasting scenarios with heterogeneous devices.
	In this paper we also study the fundamental connection between rate-distortion performance and network capacity, and propose an efficient and effective approach to train the slimmable model in a single pass.  
	
\section{Acknowledgments}

We acknowledge the support from Huawei Kirin Solution and the Spanish Government funding for projects RTI2018-102285-A-I00 and RYC2019-027020-I.

{\small
\bibliographystyle{ieee_fullname}
\bibliography{egbib}
}

\clearpage
	\renewcommand{\appendixpagename}{Supplementary material}
	\begin{appendices}
        \section{Example of $\lambda$-scheduling}
		\label{app:lambda-sched}
\begin{figure}[h]
          \centering
          \includegraphics[width=\columnwidth]{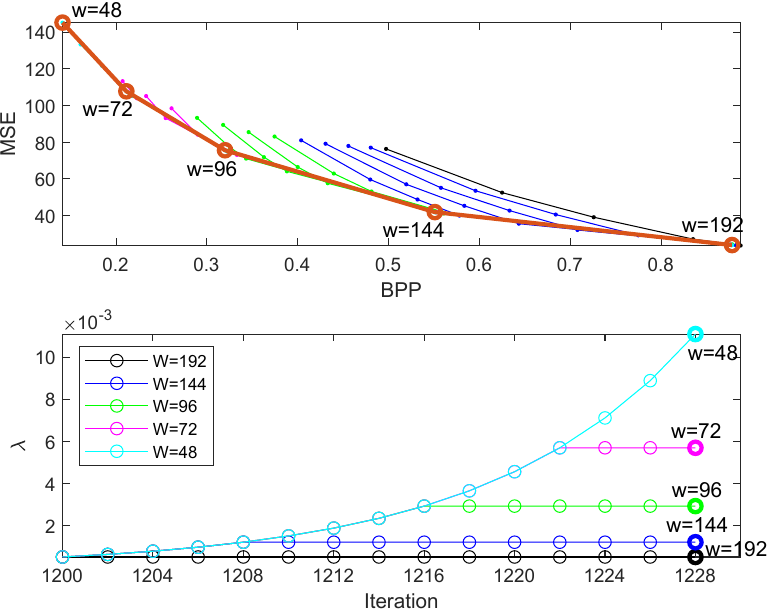}
          \caption{Evolution of the RD curve (top) and $\lambda$ during the $\lambda$-scheduling phase. Naive training shown in black (top).}
          \label{sfig:lambda-sched}
        \end{figure}

		Fig.~\ref{sfig:lambda-sched} (top) illustrates the evolution of the RD curve from the initial naive training (in black) until the end of $\lambda$-scheduling (in red). Fig.~\ref{sfig:lambda-sched} (bottom) illustrates the schedule for the $\lambda$s of the different subCAEs.		
		\section{Additional visualizations}
		\label{app:visualization}

\begin{figure*}
        \centering
         \includegraphics[height=0.95\textheight]{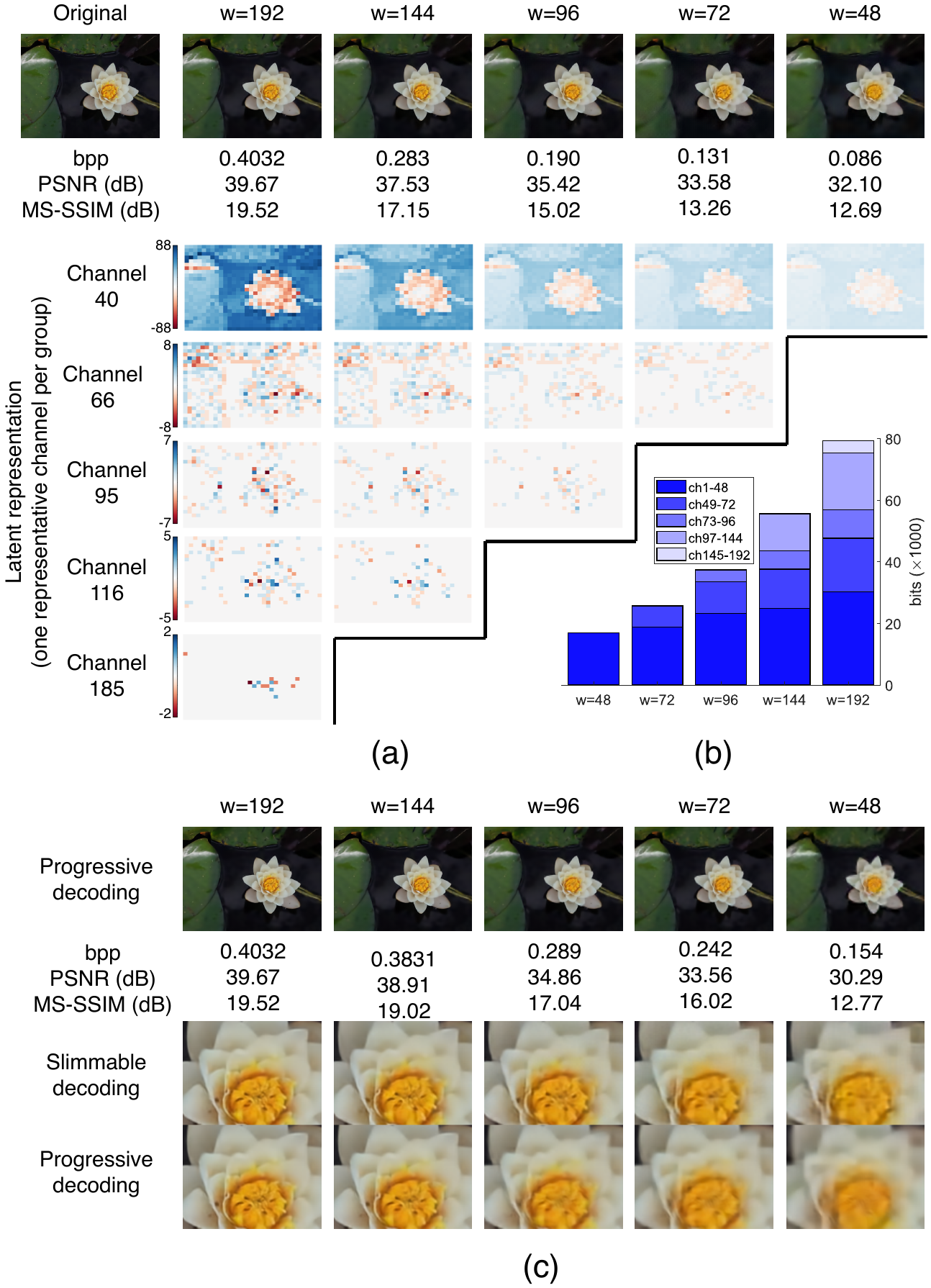}
        \caption{Illustrative example: (a, top) reconstructions using slimmable decoding, (a, bottom) selection of quantized latent maps (one per group of channels), (b) break down of bits spent in each group of channels, (c) scalable bitstream.}
        \label{latent}
        \end{figure*}

        Fig.~\ref{latent}a shows an illustrative example of the reconstructed images and (quantized) latent representations obtained by the different subCAEs. The slimmable structure of the last layer of the encoder results in a latent representation also structured in five groups of channels (i.e. 1-48, 49-72, 73-96, 97-144 and 145-192). One representative channel per group is shown. The corresponding bit allocation is shown in Fig.~\ref{latent}b. Note that smaller subCAEs also allocate fewer bits in the first group of channels.
        
        Fig.~\ref{latent}c illustrates progressive decoding when (quality) scalability is enabled, compared with the default slimmable decoding (non scalable bitstreams). In this case each group of channels of $w=192$ in Fig.~\ref{latent}b correspond to the base stream (channels 1-48) and four enhancement streams, which can be progressively combined to improve quality. However, there is a noticeable increase in both rate (explained by the bit allocation shown in Fig.~\ref{latent}b) and distortion with the quality scalable bitstream.
        
	\end{appendices}
\end{document}